# An ICT Enhanced Life Quality for the Elderly in Developing Countries: Analysis Study Applied to Sri Lanka

M.F.M FIRDHOUS[a,1] and P.M. KARUNARATNE [a,2]
<sup>a</sup>*Faculty of Information Technology, University of Moratuwa, Sri Lanka*
[1] firdhous@uom.lk
[2] pmkaru@itfac.mrt.ac.lk

**Abstract.** In the recent years, the entire world has seen a tremendous increase in the elderly population. Even though countries differ in the numerical criterion for defining the old age, the UN has agreed that the cutoff of 60+ years refers to the older population. When the percentage of older population increases in a country, the country faces new challenges in the form of economical as well as social impacts. The economic impacts are felt in the form of a shortfall in labor supply, income, household savings while an increase in the payment of retirement benefits and healthcare expenditures. On the social side, the elderly population feels continuously isolated due to the changes in the value system. The United Nations has identified five quality of life characteristics for the elderly in its 1991 resolution named the United Nations Principles for Older Persons.
In this paper the authors initially look at the trends in the increase of aged population in Asia in general and Sri Lanka in particular. Then they discuss how the quality of life characteristics can be achieved in a cost effective way through the use of Information and Communication Technology (ICT). The authors take an in depth look at the recent developments in field of ICT and its penetration in developing countries especially Sri Lanka with special emphasis to constraints and challenges in adopting ICT for the improving the quality of life of elderly.

## 1. Introduction

International Standards Organization defines quality as something that can be determined by comparing a set of inherent characteristics with a set of requirements. If those inherent characteristics meet all requirements, high or excellent quality is achieved. If those characteristics do not meet all requirements, a low or poor level of quality is achieved. The same document defines Quality Assurance (QA) as a set of activities intended to establish confidence that quality requirements will be met.[1] Avedis Donabedian in his seminal work defines the quality in healthcare as the product of two factors. One is the science and technology of healthcare and the second is the application of that science and technology in practice.[2] He suggests a three pronged approach of assessing quality of care: structure, process, and outcome. Structure includes material and human resources and organizational characteristics. Process is the set of activities that constitute health care, such as diagnosis and treatment, usually carried out by professional personnel but also by patients and family. Outcome is defined as the changes in individuals attributable to the care they received. It was also emphasized that these areas are not mutually exclusive and illustrate the cause and effect relationship between them.

## 2. Growth of Elderly Population

The World Health Organization (WHO) has defined health as the state of complete physical, mental and social well-being and not merely the absence of disease or infirmity.[3] The American Heritage® Medical Dictionary defines health care as the prevention, treatment, and management of illness and the preservation of mental and physical well-being through the services offered by the medical and allied health professions. Although there are commonly used definitions of old age, there is no general agreement on the age at which a person becomes old. The common use of a calendar age to mark the threshold of old age assumes equivalence with biological age, yet at the same time, it is generally accepted that these two are not necessarily synonymous.[4] In the United States of America, and the United Kingdom, the age of 65 was traditionally considered the beginning of the senior years.[5] At the moment, there is no United Nations standard numerical criterion, but the UN agreed cutoff is 60+ years to refer to the older population .[4]

In 1950, the elderly in Asia numbered roughly 57.6 million and accounted for no more than 4.1% of the region's population. By the middle of this century, the elderly population is projected to reach 922.7 million, and their share is expected to rise to 17.5%. Asia accounted for only 44% of the global elderly population in 1950, but by 2050, this share is projected to increase to 62%.[6]

Figure 1 shows the percantage of population aged 65 or older in Asia and Sri Lanka. Figure 2 shows the old age dependancy ratio for Asia and Sri Lanka. Both graphs show that the percentage of older people in Asia as well as Sri Lanka is rising rapidly and this is a burden on the working age younger group of people whose percantage is going to decrease relative to that of the aged population.

Table 1 shows that the aging population in Sri Lanka has been increasing rapidly during the last century indicating a fourfold increase in the ageing index from 6.4 to 24.0.

**Table 1:** Population by Age, Dependency Ratio & Ageing Index

| Year | Population by Age Group In Thousands | | | Dependency Ratio % | | | Ageing Index C/A |
|---|---|---|---|---|---|---|---|
| | 0-14 yrs A | 15-64 yrs B | 65yrs & over C | Total (A+C)/B | 0-14 yrs A/B | 65 yrs & over C/B | |
| 1901 | 1496 | 1963 | 96 | 81.12 | 76.2 | 4.9 | 6.4 |
| 1911 | 1680 | 2332 | 94 | 76.07 | 72.0 | 4.0 | 5.6 |
| 1921 | 1771 | 2619 | 108 | 71.75 | 67.6 | 4.1 | 6.1 |
| 1946 | 2478 | 3949 | 229 | 68.57 | 62.8 | 5.8 | 9.3 |
| 1953 | 3412 | 4595 | 283 | 80.41 | 74.2 | 6.2 | 8.3 |
| 1963 | 4390 | 5744 | 379 | 83.03 | 76.4 | 6.6 | 8.6 |
| 1971 | 4945 | 7206 | 539 | 76.10 | 68.6 | 7.5 | 10.9 |
| 1981 | 5227 | 8979 | 641 | 65.35 | 58.2 | 7.1 | 12.3 |
| 2001 | 4449 | 11413 | 1068 | 48.30 | 39.0 | 9.4 | 24.0 |

Source: *Department of Census and Statistics*

The aging population impacts a country in several ways. It puts a higher financial burden in the form of more retirement benefits to be paid and also the state has to shoulder a higher healthcare expenditure, in addition to these it will have an impact on the labor supply and income, household savings and consumption too.[6]

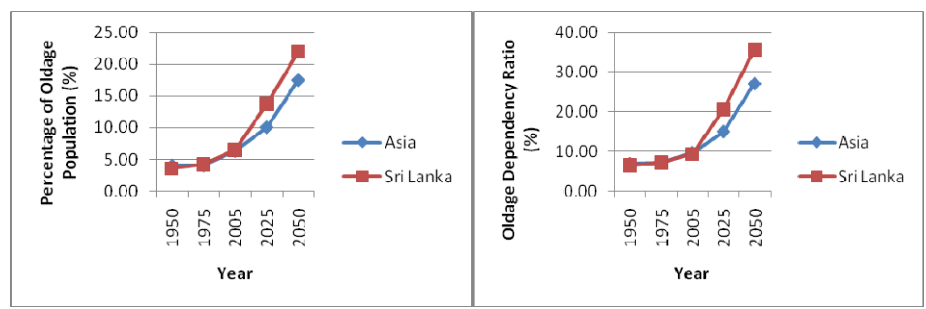

**Figure 1:** Percentage of Population Aged 65 or above     **Figure 2:** Old Age Dependency Ratio

Data Source: *UN World Population Prospects:*     Data Source: *UN World Population*
*The 2006 Revision*                                *Prospects: The 2006 Revision*

The elderly in Asia have traditionally relied on filial resources for old-age support, but the extent to which they can continue to do so has become increasingly uncertain. As extended family networks wane and more modern ideas about marriage, family, and individualism take hold, the fastest growing segment of the population will have no other recourse but to turn to public or private institutions for support.[6] As people continue to live longer, the challenge for public health is to increase the number of years a person lives free of disability, primarily by ensuring access to high quality, affordable and sustainable health and care services. Policymakers recognize, however, that modern health-care systems are not always equipped to deal with the growing ageing population, because they have been set up for acute care and expediency, rather than attending to the chronic illnesses common in older patients. Although family members continue to provide most of the long-term care of older persons, factors such as the shift from extended to nuclear families, the rise in rural-to-urban and international migration, and the increase in labor force participation rates of women have combined to increase the strain on such informal care-giving arrangements.[7]

"The United Nations Principles for Older Persons" the resolution adopted by the United Nations General Assembly (UNGA) in 1991 has identified 18 Quality of Life Characteristics for the elderly under five groups.[7,8] They are:

- independence
- participation
- care
- self-fulfillment
- dignity

In this paper, the authors look at how the quality of life characteristics identified by the United Nations General Assembly can be met in a cost effective way through the use of advanced technology.

## 3. Information and Communication Technology

Information and Communication Technology in short ICT can be defined as tools that facilitate communication and the processing and transmission of information and the sharing of knowledge by electronic means. This encompasses the full range of electronic digital and analog ICTs, from radio and television to telephones (fixed and mobile), computers, electronic-based media such as digital text and audio-video recording, and the Internet, but excludes the non-electronic technologies.[9]

ICTs offer major transformational opportunities. They can contribute to enhanced productivity, competitiveness, growth, wealth creation, poverty reduction and can spur the knowledge-based economy. ICTs provide the means by which knowledge is developed, stored, aggregated, manipulated and diffused. ICTs also enable participation in the global economy. The New Economy, the Information Society and associated transformations and opportunities reach out and engage all countries.[10] These opportunities are well known and are not just a developed country phenomenon. ICTs, particularly access to broadband Internet, are vital for developing nations as well. The ITU's Build on Broadband project is dedicated to promoting equitable, affordable broadband access to the Internet for all people, regardless of where they live or their financial circumstances.[11] In the 21st century, affordable broadband access to the Internet is becoming as vital to social and economic development as networks like transport, water and power. Broadband access – and the next generation broadband network infrastructure which underpins it – is a key enabler for economic and social growth. Broadband changes everything. It enables not just great new enabling applications, such as VoIP and IPTV, but also the delivery of essential services – from e-health to e-education to e-commerce to e-government. And broadband is helping us make great progress towards meeting the Millennium Development Goals – and improving the quality of life for countless people around the world.[12]

ICTs can have an impact on everyday lives and on general economic activity, but the opportunities only materialize fully to the extent that the regulatory framework, as implemented, supports and fosters both investment in and widespread diffusion of ICTs. Absent these conditions, the full promise of ICTs is unrealized. ICTs offer the prospects of rapid advancements, but if appropriate conditions are not in place, the outcome can be a rapid slide down the digital divide. And although the digital divide is narrowing, particularly due to the rise of Internet-enabled mobile phones and applications, a new broadband divide is growing that governments need to address.[10,13]

There are some stunning successes, particularly with regard to mobile services. In 2002, the total number of mobile subscribers in the world surpassed that of fixed customers. Between 2004 and 2009, mobile phone subscriptions worldwide grew from nearly 1.8 billion to an estimated 4.6 billon, translating into a growth in mobile penetration from less than 28 percent to 67 percent.[10]

The Asia-Pacific region is the largest mobile market in the world, and by 2013, Asia is expected to have almost three billion mobile subscribers. In 2008, China alone had 634 million mobile subscribers, which far exceeded the combined number of mobile subscribers in Japan and the United States at 110 million and 270.5 million subscribers, respectively. Sub-Saharan Africa had a mobile penetration of rate of 32 subscribers per 100 people in 2008, this translated into over 246 million mobile customers.[10,14]

Mobile phone handsets are now turning into smart-phones equipped with digital cameras, Internet-enabled video, pre-installed social networking applications such as Facebook and music juke box payment terminals. These new functionalities are transformational. For example, as digital cameras, mobile devices provide benefits such as instant news gathering. Their Internet-enabled video, access to social networks and music capability brings them into the realm of media, copyright and Internet governance. As a component of the banking system, the mobile network can provide services where the financial network is weak. ICTs have significantly impacted business operations where a large number of new, non-OECD countries have successfully entered the market. This is particularly the case for software and IT Enabled Services. Market entry is partly explained by the "death of distance" or the dramatic fall in the costs of international connectivity. The latest manifestation is the proliferation of broadband access networks. Broadband can carry huge quantities of data, at very high speeds. Although postal and courier services can deliver large quantities of data (e.g., a truckload of CDs), they fail the speed test. In the broadband world, large volumes of data can be moved almost instantaneously to widely dispersed locations at low cost. Through the application of ICTs, many services once considered non-tradable are now tradable, such as back-office functions including the management of employee benefits or dental records. "Out-sourcing" and/or "business process off-shoring" (BPO) have seen massive increases, amounting to a total addressable market estimated at US$ 300 billion, of which US$ 100 billion will be off-shored by 2010. In the BPO market, India is a tremendous success story. It has become the dominant player in the BPO market. Growth in India's BPO exports were 44.5 percent in 2005 and employment in the sector increased from 42,000 jobs in 2002 to an estimated 470,000 in 2006. Other countries like the Philippines, Brazil, Romania and Ireland have also been particularly successful in attracting investment and creating employment from BPO-related activities. The use of ICTs in e-government services is also transforming citizens' interactions with the public sector by improving efficiency, effectiveness and accountability of governments. In India, for example, a comparison of manual and e-government services found that computerized services substantially increased cost-savings and access to services.[10]

### 4. ICT Penetration in Sri Lanka

Sri Lanka is not far behind other countries in the Asian region in terms of ICT penetration. The ICT penetration can be looked at as the number of users under different categories. Table 2 shows the percentage of households having computers sector wise as well as province wise. From this data, it can be seen that the household computer availability in Urban sector (23.6%), where a computer is available in one out of every four households is much higher than the Rural sector (9.2%) and the Estate sector (3.1%) in 2009. Also the western province has the highest computers ownership (20.7%) compared to other provinces.

**Table 2:** Percentage of Computer Owned Households by Sector and Province 2004, 2006/07 and 2009

|  | Desktop (%) | | | Desktop or Laptop (%) |
|---|---|---|---|---|
|  | 2004 | 2006/07 | 2009 | 2009 |
| **Overall** | 3.8 | 8.2 | 10.6 | 11.4 |
| **Sector** | | | | |
| Urban | 10.5 | 17.8 | 23.6 | 26.3 |
| Rural | 3.1 | 6.9 | 9.2 | 9.8 |
| Estate | 0.3 | 1.1 | 3.1 | 3.3 |
| **Province** | | | | |
| Western | 8.4 | 16.4 | 19.0 | 20.7 |
| Central | 3.3 | 6.7 | 9.7 | 10.4 |
| Southern | 2.2 | 4.9 | 6.6 | 7.2 |
| Eastern | 1.2 | 3.7 | 5.8 | 5.9 |
| North-western | 3.1 | 4.8 | 6.9 | 7.1 |
| North-central | 1.4 | 2.7 | 6.1 | 6.8 |
| Uva | 0.4 | 2.7 | 4.6 | 4.9 |
| Sabaragamuwa | 2.0 | 3.3 | 7.3 | 7.5 |

Data Source: *Dept. of Census and Statistics of Sri Lanka*

Table 3 shows the percentage computer awareness and computer literacy among different sectors of the Sri Lankan households. It can be seen that 44 percent of the population in the age group of 5–69 years aware about computers in 2009 in Sri Lanka and it was 37 percent in 2006/07. There are significant differences in computer awareness across the sectors. The highest computer awareness is reported from the Urban sector (60.0%) households and the lowest by the Estate sector (15.8%) households. Among the provinces the highest computer awareness is reported by the Western province (51%) and the lowest is from the Uva province (29%). Similar pattern can be seen for computer literacy as well. Computer literacy reported in 2009 in Sri Lanka is 20.3 percent. The highest (31.1%) computer literacy is reported from the Urban sector households and the lowest (8.4%) is reported among the Estate sector household population. In Urban areas, almost one out of every three persons is a computer literate person. Among the provinces the highest level of computer literacy is also reported from the Western province (28%) and the least level is in the Eastern province (13%).

**Table 3:** Computer Awareness and Computer Literacy of Household Population (aged 5 - 69) by Sector and Province

|  | Computer Awareness (%) |  | Computer Literacy (%) |  |
| --- | --- | --- | --- | --- |
|  | 2006/07 | 2009 | 2006/07 | 2009 |
| **Overall** | 37.1 | 43.8 | 16.1 | 20.3 |
| **Sector** |  |  |  |  |
| Urban | 47.4 | 60.0 | 25.1 | 31.1 |
| Rural | 36.9 | 43.0 | 15.1 | 19.3 |
| Estate | 10.3 | 15.8 | 4.3 | 8.4 |
| **Province** |  |  |  |  |
| Western | 47.9 | 50.7 | 23.2 | 27.7 |
| Central | 31.0 | 34.8 | 14.8 | 18.0 |
| Southern | 43.2 | 45.0 | 15.6 | 19.8 |
| Eastern | 31.5 | 46.6 | 11.4 | 12.9 |
| North-western | 31.8 | 42.1 | 12.6 | 16.5 |
| North-central | 27.5 | 40.4 | 8.9 | 14.1 |
| Uva | 22.3 | 29.3 | 9.9 | 14.7 |
| Sabaragamuwa | 30.2 | 44.6 | 12.3 | 19.1 |

Data Source: *Dept. of Census and Statistics of Sri Lanka*

Figure 3 shows the computer literacy of population in 2009. From Figure 3, it can be seen that only about 3 percent of the 60-69 age group or the elderly population are computer literate. Figure 4 shows the number of Internet and email users in Sri Lanka over the years. This sector has seen a steady growth from the time Internet was introduced to Sri Lanka in mid 90s and reached 250,000 subscribers in 2009.

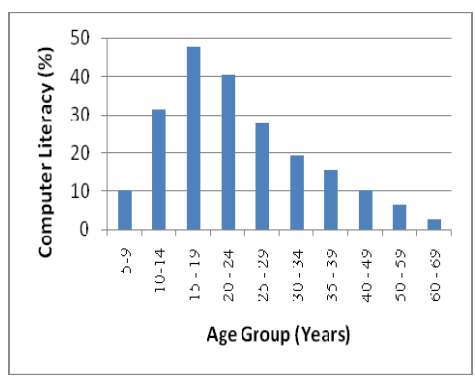

**Figure 3:** Computer Literacy of Population Aged 5 – 69 Years - 2009

Data Source: *Dept. of Census and Statistics of Sri Lanka*

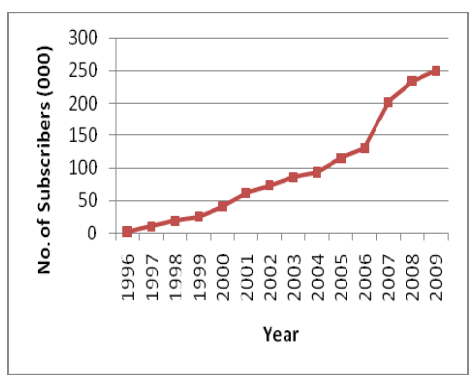

**Figure 4:** Number of Internet and Email Users over the Years

Data Source: *Telecommunications Regulatory Commission of Sri Lanka*

Figure 5 shows percentage household population by age group who has used Internet and email during the last 12 months in 2009. From Figure 5, it can be seen that among age group of 60-69 years, the Internet and email usage is very low. Figure 6 shows the number of mobile phone users in Sri Lanka over the years. There is a rapid increase in mobile phone users after the year 2000. This increase can be attributed to the low cost of mobile phones and lowering tariff due to high competition among the operators. During this period, mobile operators started introducing new value added services like SMS, video message service, mobile internet, news alerts etc., in addition to the basic voice services. This became possible due to the digitalization of the entire mobile infrastructure along with the modernization of the network.

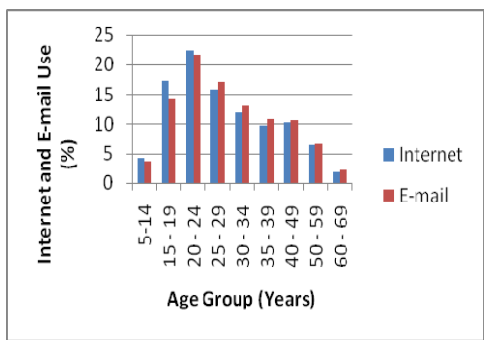

**Figure 5:** Households Population Aged 5–69 Years Who Used Internet and E-mail at least Once During the Past 12 Months

Data Source: *Dept. of Census and Statistics of Sri Lanka*

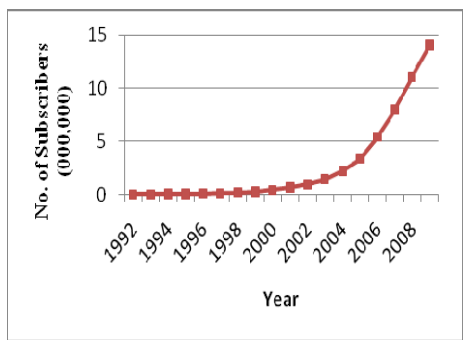

**Figure 6:** Number of mobile phone users over the Years

Data Source: *Telecommunications Regulatory Commission of Sri Lanka*

The recent literature has identified connectivity, content, capacity, community, commerce, culture, cooperation and capital as the factors that can inhibit the introduction of ICTs in the health sector.[9] These eight factors can be called the 8Cs of ICT penetration of a country. The situation in Sri Lanka can be analyzed in the light of these 8Cs.

Connectivity deals with the policies, regulations and the availability of enabling technology such as the ownership of equipment and the available infrastructure. Data on Sri Lanka shows that there is high penetration of mobile devices along with well advanced infrastructure with 3G and 3.5G technologies. Mobile networks with 3G or 3.5G technology is capable of carrying voice, high speed data and video streams.

Content factors include the lack of local content creation, the language used and the relevance of content to the local situation. In the Sri Lankan case, the content factor is present but the content relevant to local requirements can be developed with relative ease as the government has already taken an initiative for the localization of ICT that includes development of computer applications with local language interfaces and content in the local languages.

Capacity deals with two important issues. One is the capacity of end users to use ICT effectively and the other one is the skilled ICT work force for the implementation and management of the ICT infrastructure. Since Sri Lanka has a skilled work force along with a high literacy rate, meeting the capacity requirement will not be a major problem provided the content is already available in the local languages.

Community refers to the users of ICT. Almost every Sri Lankan is familiar with some sort of ICT. Also Sri Lankans are very receptive to new technologies. This can be seen from the fact how fast new technologies like the Internet or mobile phone access grew in the past.

Commerce factor deals with the degree to which it is possible to develop effective national or domestic Internet economies that promote online transactional capabilities that will be beneficial for consumers, and for businesses. The effect of this factor is yet to be seen in Sri Lanka as the other supporting technologies like payments methods, delivery of goods and services purchased online are yet to be developed.

Culture deals with two aspects. One is the appropriate and relevant content and the other aspect is the need to examine and challenge the cultural inhibitions and barriers within society and institutions that prevent effective use of ICTs. In the case of Sri Lanka, these factors have already been discussed under content factor and community factor.

Cooperation deals with the interaction of different stakeholders for the effective use of ICT. Dzenowagis has identified six major groups as the main stakeholders in effective use of healthcare ICT. They are namely; citizens (including patients), professionals, hospitals and academia, health-related businesses, governments and international agencies.[15] Sri Lankan case is yet to be analyzed on this issue as currently there is no such application that includes every one of these parties.

Capital is the total investment required to implement ICT related infrastructure, applications and development of required skills. Since, Sri Lanka is a low income country this can be a real problem unless some kind of assistance is not available from international donors. But Sri Lanka already processes a well developed ICT infrastructure in the form of the mobile and fixed communication network and also the Sri Lankan users can be trained on any new technology with relative ease and low cost.

Considering the constraints and challenges in the form of 8Cs above and the data on existing infrastructure and penetration, the most promising technology to be used immediately is the mobile phone technology. As discussed in Section III, the mobile phones come with a lot of additional features like digital cameras, Internet enabled and preinstalled Internet based applications, the mobile phone can be used for mobilizing any new application very easily. Since the advanced mobile networks like 3G or 3.5G have voice, data and video capabilities, no new investment is not required when new internet based applications are introduced on the mobile platform. So the mobile phones and the mobile network can be used to launch new application to meet the requirements of the elderly population in Sri Lanka.

### 5. Empowering the Aged Through ICT

Many of the Quality of Life Characteristics identified by the UNGA are centered around how to make the elderly inclusive in the society. That is they should be allowed to participate and contribute to the society in whatever way possible and made to feel that they are an integrated part of the society. The characteristics 2,3,4,5,7,8,9,11,12,13,14,15 and 16 deal with the empowerment of the elderly so that they can contribute to the society rather than depending on the society for their survival. The characteristics in detail are as follows:

2. Older persons should have the opportunity to work or to have access to other income generating opportunities.
3. Older persons should be able to participate in determining when and what pace withdrawal from the labour force takes place.

4. Older persons should have access to appropriate educational and training opportunities.
5. Older persons should be able to live in environments that are safe and adaptable to personal preferences and changing capacities.
6. Older persons should remain integrated in society, participate actively in the formulation and implementation of policies that directly affect their well being and share their knowledge and skills with younger generations.
7. Older persons should be able to seek and develop opportunities for service to the community and to serve as volunteers in positions appropriate to their interests and capabilities.
8. Older persons should be able to form movements or association of older persons.
9. Older persons should have access to health care to help them to maintain or regain the optimum level of physical, mental and emotional well-being and to prevent or delay the onset of illness.
10. Older persons should have access to social and legal services to enhance their autonomy, protection and care.
11. Older persons should be able to utilize appropriate levels of institutional care providing protection, rehabilitation and social and mental stimulation in a humane and secure environment.
12. Older persons should be able to enjoy human rights and fundamental freedoms when residing in any shelter, care or treatment facility, including full respect for their dignity, beliefs, needs and privacy and for the right to make decisions about their care and the quality of their lives.
13. Older persons should be able to pursue opportunities for the full development of their potential.
14. Older persons should have access to the educational, cultural, spiritual and recreational resources of society.

All these characteristics can be achieved by empowering the elderly in such a manner that they could continue to contribute to the society in a positive manner as long as they would want to. One of the main issues with elderly is they are compulsorily made to retire from their professional lives due to government regulations. Even though the age of retirement varies depending on the profession this happens to every person other than those who are self employed. The self employed may retire from their professional life due to ill health or other disability. Except a selected few, who are recalled to the working life as advisors or consultants by the government, private or nongovernmental organizations, all the others say goodbye to their profession totally. Most of these elderly posses a wealth of knowledge and information gained throughout their career life. This knowledge can be used for the betterment of the society and the future generation if tapped and used properly. With little training on modern technologies and modifying the technologies to suit the needs and the capabilities of these elders they can again be made useful members of the society.

The elders can be made to contribute their knowledge to the society by creating an elders' portal. This portal must be created in such a manner that the older persons with limited ICT literacy, limited English language capability, and limited physical movement will be able to access and use it. This portal must include services like knowledge portals in different knowledge areas, social networking facility, entertainment sites, links to eGovernment services etc.

The main requirements that need to be met when setting up such portals are:

- must use existing technology in order to avoid the cost of setting up of new infrastructure and the delay of implementation.
- must be user friendly in such a manner that elderly people must be able to access the site with relative ease.
- must support have native language support.

*5.1. Knowledge Portals*

Knowledge portals can be setup in different knowledge areas. For example, an agriculture knowledge portal can be setup for the elders who were engaged in agriculture and now retired. Using this agriculture portal a farmer who has a long experience in rice farming in the dry zone will be able to help the young farmers with his wealth of knowledge. But for an elderly person to use this portal, the technology must be user friendly from his point of view. The first requirement is, the portal must be in either Tamil or Sinhala. Since the computer literacy of the elders is very minimal especially in the rural areas, the most probable technology would be mobile technology as it would be possible to train these people with minimum effort on this. The other advantage is the penetration of the mobile communication industry is very high compared to the computer industry in Sri Lanka while the cost of mobile phones and usage is very low compared to that of computers. Mobile phone has a limitation compared to the computer. Mobile phones have a keypad with only 12 keys. Typing anything on this keypad other than a number is a cumbersome operation as multiple keystrokes or a combination of keystrokes will have to be used to type a single character. This problem becomes more acute as Sinhala or Tamil is used instead of English as the number of characters in these two languages is much higher than that of the English language. Hence the most probable way to get the elderly involved in this exercise to use voice with minimal text input. Hence, the voice mail facility or audio posting of questions and answers must be available. Many elders of today were educated in the English medium. Another similar knowledge portal can be setup where the language capabilities of these elders can be used effectively coach the younger generation.

*5.2. Social Networking*

Social networks like facebook, linkedin and twitter are very popular among youth. An elderly friendly social network in the native language will provide the elders to meet associate and spend time with people of their choice online. If the portal supports social networking facility with voice and video capability, they would be able to use this facility from their home on a smart phone.

*5.3. Entertainment Facilities*

Online entertainment facilities would provide them with recreational facilities. These entertainments could be online games, music, video or any other from which the elders would be able access and enjoy with relative ease.

*5.4. Access to eGovernment and Other Services*

This portal can be linked to other public and private services where they would be able access with relative ease.

*5.5. Additional Services*

Since the mobile phones of today are actually smart phones, the facilities in these phones can be used to provide the elders with protection and care. If the portal can be used linked to an online monitoring site, the camera in the phone can be used to monitor the elders and alert the relevant parties in case of emergencies. This portal can be used to remind the elders to take their medicine on time through an alert service.

Figure 8 shows a typical interface for this portal on a smart phone. It can be seen from this figure, the interface provides the capability to select the language in which a user wants to use the portal. Also the interface provided is very simple in terms of information provided and the user interaction required. The user needs to press a button to select his or her choice. This is far short of a multimedia rich web portal but an older user would be able to use this interface very efficiently compared to a multimedia interface where the choice needs to be made by pointing and clicking using a pointing device such as a mouse.

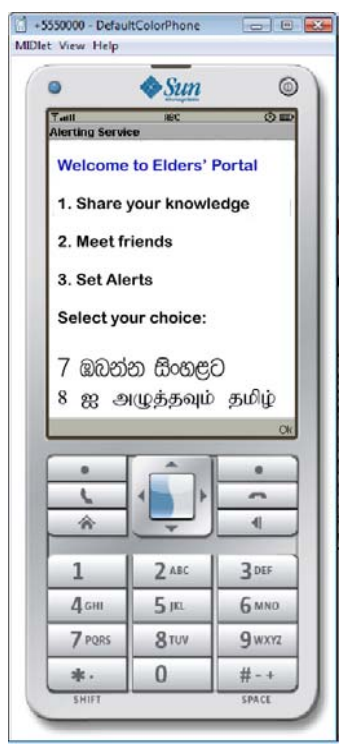

**Figure 8:** Interface for an Elders' Portal

## References

[1] "*Definition of an Older or Elderly Person*", http://www.who.int/healthinfo/survey/ageingdefnolder/en/index.html [Accessed on 26th July 2010]
[2] Donabedian A, "*An Introduction to Quality Assurance in Healthcare*", Oxford University Press, 2003.
[3] Dzenowagis J, "*Connecting for Health: Global Vision, Local Insight*" World Health Organization, Geneva, 2005.
[4] *General Assembly resolution 46/91 of 16 December 1991*, annex: "United Nations Principles for Older Persons: to add life to the years that have been added to life", United Nations, New York, 1991.
[5] Heeks R, "*The ICT4D 2.0 Manifesto: Where Next for ICTs and International Development?*", Paper No. 42, Development Informatics working Paper Series, Development Informatics Group, Institute for Development Policy and Management, University of Manchester, UK, 2009.
[6] infoDev, "*Improving Health, Connecting People: the Role of ICTs in the Health Sector of Developing Countries - A Framework Paper*", Information for Development Program, World Bank, USA, 2006.
[7] InfoDev, ITU, "*Regulating the Telecommunications/ ICT Sector: Overview*", ICT Regulatory Toolkit, http://www.ictregulationtoolkit.org [Accessed on 25th July 2010]
[8] ITU, "*Build on Broadband*", http://www.itu.int/en/broadband/Pages/default.aspx [Accessed on 25th July 2010]
[9] ITU, "*ICT Statistics Database*" http://www.itu.int/ITU-D/icteye/Indicators/Indicators.aspx# [Accessed on 25th July 2010]
[10] "*ISO9000, 9001, and 9004 Quality Management Definitions*", http://www.praxiom.com/iso-definition.htm#Quality [Accessed on 26th July 2010]
[11] Menon J, Melendez-Nakamura A, "*Aging in Asia: Trends, Impacts and Responses*", ADB Working Paper Series on Regional Economic Integration No. 25, Asian Development Bank, Manila, Philippines, 2009.
[12] Touré HI, "*The Importance of ICTs and Broadband as Bital Enablers for Social and Economic Development*", ITU Secretary-General, Conferencia Magistral, Santo Domingo, Dominican Republic, 2009.
[13] Wikipedia, "*Old Age*", http://en.wikipedia.org/wiki/Old_age [Accessed on 26th July 2010]
[14] *WHO: Preamble to the Constitution of the World Health Organization as adopted by the International Health Conference*. International Health Conference, New York, 19–22 June, 1946, 2:100.
[15] Zelenev S, "*The Madrid Plan: A Comprehensive Agenda for an Ageing World*".